\documentclass{rspublic}
\input epsf.tex
 
\def\hh{\rm H_{2}}
\def\hxp{\rm HX^{+}}
\def\xyp{\rm XY^{+}}
 
\def\hhhp{\rm H_{3}^{+}}
\def\hcop{HCO$^{+}$}
\def\hhp{\rm H_{2}^{+}}
\def\oo{O$_{2}$}
\def\um{$\mu$m}

\begin{document}

\title[H$_{3}^{+}$~between the stars]{H$_{3}^{+}$ between the stars} 

\author[T. R. Geballe]{Thomas R. Geballe}

\affiliation{Gemini Observatory, 670 N. A'ohoku Place, Hilo, HI 96720,USA}

\label{firstpage}

\maketitle

\begin{abstract}{infrared spectroscopy, interstellar clouds, interstellar 
molecules} 

The presence of $\hhhp$ in the interstellar medium was forecast almost four
decades ago. Almost three decades ago it was asserted that its reactions
with neutral molecular and atomic species directly lead to the production
of many of the interstellar molecules that have been discovered by radio
and infrared astronomers. With the recent detection of $\hhhp$ in
interstellar space, astronomers finally have direct confirmation of $\hhhp$
as the foundation of ion-molecule interstellar chemistry. Although many
questions remain to be answered, it is clear that $\hhhp$ is a unique tool
for understanding the properties of interstellar clouds.

\end{abstract}

\section{Dark Clouds and the Role of $\hhhp$ }

Half a century ago astronomers were just becoming aware that
interstellar space contains considerable quantities of hydrogen, in both
atomic and molecular form. Almost four decades ago Martin \textit{et al.}
(1961) pointed out to astronomers that, where interstellar $\hh$ is
ionized, $\hhhp$ is produced rapidly as a result of the reaction,
  \begin{equation} 
  \hhp~+~\hh~\rightarrow~\hhhp~+~H,
  \end{equation} 
\noindent by which $\hhhp$ is created in abundance in laboratory hydrogen
plasmas.  Molecular hydrogen is the dominant hydrogenic species in dark
clouds, where dust particles prevent the penetration of ultraviolet
radiation. Solomon and Werner (1971) recognized that within dark clouds 
cosmic ray ionisation of $\hh$,
  \begin{equation}
  \hh~+~c.r.~\rightarrow~\hhp~+~e^{-}~+~c.r.,
  \end{equation} \noindent is the principal means of production of $\hhp$
and, through it, $\hhhp$. The flux of cosmic rays is such that an
individual $\hh$ molecule is ionised roughly once per billion years. In a
cloud of density 10$^{4}~$cm$^{-3}$ each newly created $\hhp$ ion survives
for about one day before undergoing reaction 1.1.

At about the same time discoveries by radio and millimetre wave
spectroscopists of a variety of simple molecules (including free radicals)
in dark clouds (e.g., Rank \textit{et al.} 1971) were being reported. The
discoveries clearly implied the existence of an active chemistry in these
cold and rarified regions. The proposal by Klemperer (1970) that an
unidentified intense line at a wavelength of 3.4~mm, originally referred
to as "X-ogen", was emitted by \hcop\ (later confirmed when the
corresponding line of the $^{13}$C isotope of that molecular ion was
detected by Snyder \textit{et al.} 1976) suggested that gas phase
ion-neutral reactions, which have no activation energy barriers, could be
important in dark clouds.

In 1973 Watson and, independently, Herbst \& Klemperer incorporated the
forgoing ideas into detailed models for the gas-phase chemistry of dark
clouds, proposing networks of ion-molecule reactions as the means of
production for the simple molecules observed in dark clouds. Herbst \&
Klemperer approximately reproduced the observed abundances of some
of these molecules. These two papers revealed for the first time the
fundamental importance of $\hhhp$. In their models as well as in those of a
multitude of related papers that have followed, $\hhhp$ is the principal
initiator of reaction chains via the generic reaction,
  \begin{equation}
  \hhhp~+~X~\rightarrow~\hxp~+~\hh,
  \end{equation}
\noindent where X is almost any constituent of the cloud (He,
\oo\, and N are exceptions). The product ion $\hxp$ then combines
with other species through
  \begin{equation}
  \hxp~+~Y~\rightarrow~\xyp~+~H,
  \end{equation} \noindent and so on, creating networks of reactions, as
first detailed by Watson and by Herbst \& Klemperer. Later papers enlarged
and refined the early models for dark clouds (e.g., Lee \textit{et al.}
1996) and extended and adapted the basic ideas to diffuse clouds (van
Dishoeck \& Black 1986). Reactions 1.3 serve as sinks for $\hhhp$,
severely reducing its steady-state abundance, because rate coefficients
of $\hhhp$ with the most abundant species are large.  Dissociative
recombination on electrons (reaction 1.3 with X~=~e$^{-}$) has a very large
coefficient and is an important sink where electrons densities are
sufficiently high.

\section{Search strategies and early searches}

Although compelling evidence for the importance of ion-molecule chemistry
in the insterstellar medium has abounded since the 1970's, the ultimate
test of its significance would be the direct detection of $\hhhp$ and the
determination of its abundance. To detect $\hhhp$ requires spectroscopic
measurements, but in which band and at what wavelength? $\hhhp$ has no
well-bound excited electronic states, and hence no ultraviolet or visible
line spectrum. Likewise, its lack of a permanent dipole moment prohibits a
pure rotational spectrum, which would occur at far infrared and
sub-millimetre wavelengths. The symmetric $\nu_{1}$ vibration does not
induce a dipole moment and thus has no associated vibration-rotation
transitions. However, the asymmetric $\nu_{2}$ vibration does induce a
dipole moment. Following the laboratory measurements of the fundamental
vibration-rotation band by Oka (1980), the $\nu_{2}$ band could be used to
search for $\hhhp$. In view of the expected weakness of the $\hhhp$ lines
it is fortuitous that the $\nu_{2}$ fundamental near 4~\um\ and first
overtone near 2~\um\ (which is used for studies of planetary ionospheres)
do not coincide closely with bands of astrophysically abundant molecules
and, in addition, occur at infrared wavelengths which are for the most
part accessible to ground-based telescopes.

\begin{figure}
\epsfxsize=7.5cm
\centerline{\epsfbox{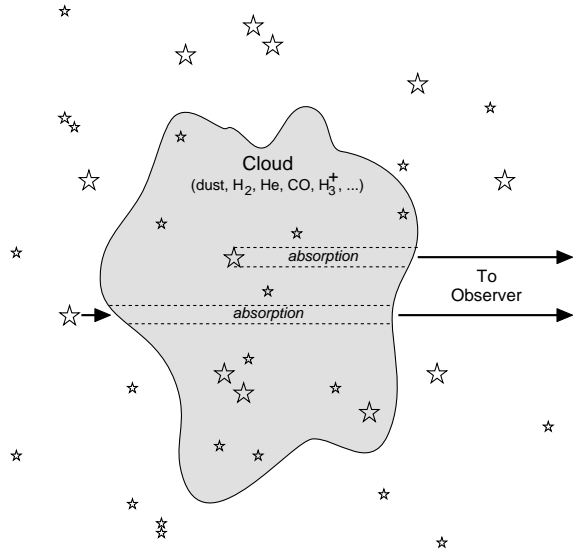}}
\caption{Absorption spectroscopy of dark clouds.}
\end{figure}

Quiescent dark clouds are the most obvious sites to search for
interstellar $\hhhp$. Since these clouds are usually very cold
(typically 10--50~K), only the lowest rotational levels of the ground
vibrational state of the molecule are populated and one is required to
search for the vibration-rotation lines associated with those levels, in
absorption against the continua of stars or protostars either embedded in
the clouds or situated behind them (figure~1). Six lines are potential
targets; four from the lowest (J=1, K=1) level of para-$\hhhp$ and two
from the lowest (J=1, K=0) level of ortho-$\hhhp$, which is 32.9~K higher
(figure~2). Note that the J=0, K=0 level is forbidden by the
Pauli exclusion principle. In a dark cloud the ortho/para ratio is
thermalised by proton hops and transfers of hydrogen atoms between $\hhhp$
and its most frequent collision partner, $\hh$ (reaction 1.3 with
X=$\hh$). The ground (para) state is more highly populated than the lowest
ortho state at temperatures less than T~$\approx$~50~K. At 50~K
molecules in the next lowest level (2,2) constitute less than five percent
of the total.
\begin{figure}
\epsfxsize=7.0cm
\centerline{\epsfbox{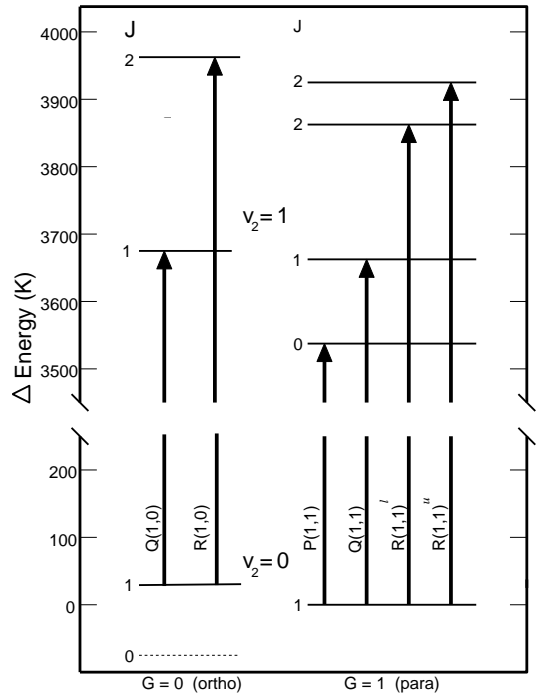}}
\caption{Vibration-rotation transitions from the lowest ortho and para states 
of $\hhhp$.}
\end{figure}
Even with essentially all of the $\hhhp$ in the two lowest energy levels,
the narrow absorption lines from those levels are expected to be very weak
because of the miniscule steady state abundance of $\hhhp$. Thus, detection
of $\hhhp$ requires the use of sensitive high resolution infrared
spectrometers, large telescopes, bright, yet highly obscured astronomical
sources of infrared continuum, and careful attention to both wavelength
calibration and the removal of atmospheric and instrumental spectral
features.

The possibility of detecting interstellar $\hhhp$ lines in emission also
should be considered. Detection of weak emission lines often is more
straightforward than detection of absorption lines, because a source of
background continuum radiation is not required and emission from a much
larger solid angle of cloud or nebula can be observed. However, in order to
detect line emission one must find environments for which not only does
$\hhhp$ exist, but also a significant fraction of it is vibrationally
excited. Within some clouds shock-excitation results from the interaction
of high velocity winds, from embedded protostars or from supernovae ejecta,
with the ambient gas. In the interaction zone collisional vibrational
excitation of $\hhhp$ and subsequent line emission should occur, as they do
in the case of $\hh$.  Conditions in planetary nebulae which are ejecting
extensive circumstellar molecular envelopes also can result in significant
vibrational excitation of $\hhhp$. However, in both of these environments
the columns of hot $\hhhp$ are very short compared to the dimensions of the
cloud or nebula, making detection difficult.

In the 1980's and the first half of the 1990's several attempts were made
to detect $\hhhp$ in a variety of interstellar environments. All of these
failed. During this period, however, the resolutions and sensitivities of
infrared spectrometers improved, largely due to the advent of
2-dimensional arrays of infrared detectors. Telescope pointing and
tracking accuracies and image sharpness also were considerably enhanced.
Each of these improvements, along with with the experience gained from the
early searches, contributed to the eventual detection of $\hhhp$.

\section{Detection in dark clouds}

The first detections of $\hhhp$ in interstellar space (Geballe \& Oka 1996)
were made toward the bright infrared sources W33A and GL2136. These objects
are high mass protostars still located deep inside their natal clouds which
were the targets of the search. The initial detections, obtained on 29
April 1996 at the United Kingdom Infrared Telescope (UKIRT) on Mauna Kea,
were decisively confirmed on July 15 of that year. Both nights'
observations utilised UKIRT's superb infrared spectrometer CGS4 (Mountain
\textit{et al.} 1990), which can obtain high resolution spectra in narrow
wavelength intervals, and focussed on the closely spaced pair of ortho and
para lines near 3.67~\um. The detected lines are only 1-2 percent deep,
much weaker than nearby atmospheric absorption lines of methane, and can
barely be discerned in the unratioed spectra (figure~3). Observing from
high and dry Mauna Kea was one key to the successful detection of this line
pair, as at lower altitude sites the telluric methane lines are stronger
and blend with nearby lines of water vapor to make the crucial wavelengths
nearly opaque. A second key was to repeat the observations at a later date,
using the change in the earth's orbital velocity to change the
Doppler-shift of the astronomical lines relative to the telluric lines
(figure 4). The correct wavelength shift between April and July was
observed, a convincing demonstration of the reality of the detection.

\begin{figure}
\epsfxsize=11cm
\centerline{\epsfbox{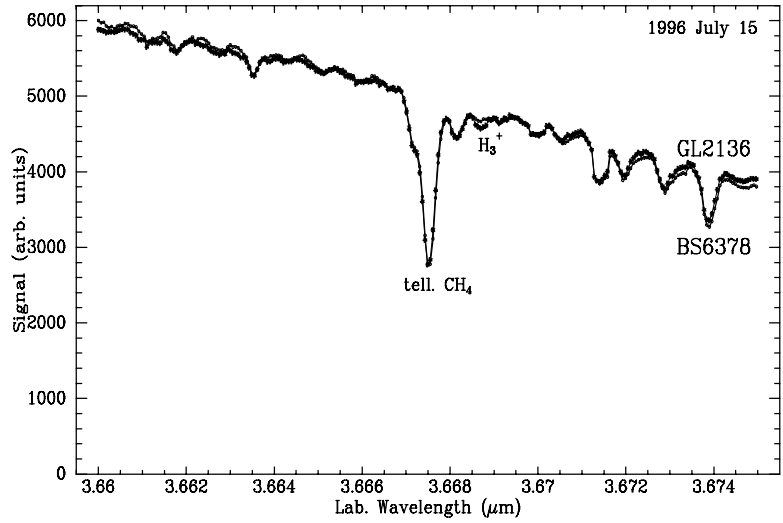}}
\caption{Raw spectra of GL2136 (thick line) and the calibration star, 
BS6378, in a spectral interval containing the $\hhhp$ ortho-para doublet,
whose location is indicated.}
\end{figure}

\begin{figure}
\epsfxsize=11cm
\centerline{\epsfbox{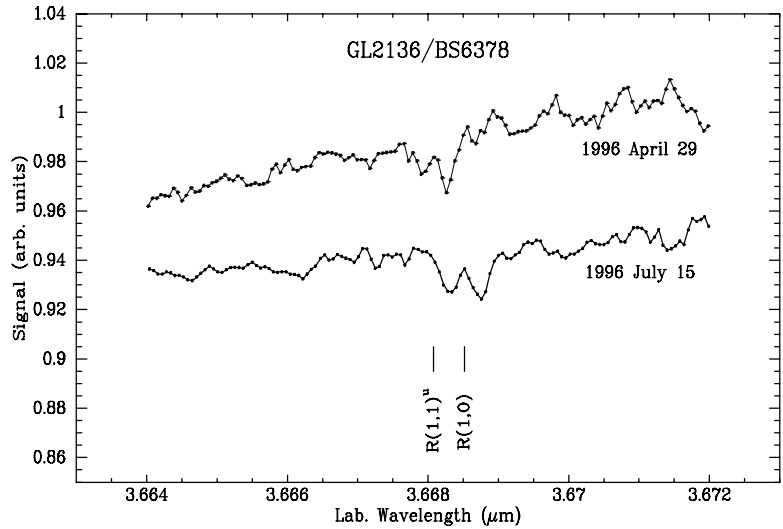}}
\caption{Ratioed spectra of GL2136 on two dates in 1996. The rest
wavelengths of the $\hhhp$ R(1,1)$^{u}$ (para) and R(1,0) (ortho) lines 
are indicated. The resolution is 15~km~s$^{-1}$.}
\end{figure}

The measurement of lines of both ortho and para $\hhhp$ in a cold dark
cloud allows the cloud temperature and the $\hhhp$ column density,
N($\hhhp$), to be determined directly. For the clouds in which W33A and
GL~2136 are situated mean temperatures of approximately 35~K and $\hhhp$
column densities of 6~$\times$~10$^{14}$~cm$^{-2}$ and
4~$\times$~10$^{14}$~cm$^{-2}$, respectively, were found (Geballe \& Oka
1996). Molecular hydrogen, the dominant constituent of dark clouds, has not
yet been detected toward W33A and GL2136. However, assuming the standard
dust-to-gas ratio found in the interstellar medium, estimates of the
N($\hh$) can be obtained from the depth of the 9.7~$\mu$m silicate dust
absorption observed toward these objects. These estimates yield values
for N($\hhhp$)/N($\hh$) of 2~$\times$~10$^{-9}$ in each of these clouds.
This compares to values of $\sim$10$^{-4}$ for the most abundant
non-hydrogenic molecule, CO, in dark clouds. $\hhhp$ indeed is a remarkably
rare constituent of these dark clouds.

The detections in W33A and GL2136 have prompted searches for $\hhhp$ in
many additional dark clouds, with detections reported in several of them
(McCall \textit{et al.} 1999; Kulesa \textit{et al.} 1999).  In all but
one case, the line strengths and derived column densities are comparable
to those found toward W33A and GL2136. With larger telescopes and improved
spectrometers now coming into use it is likely that the number of
detections will increase considerably in the next few years.

\section{Testing the ion-molecule model}

In the ion-molecule model of dark cloud chemistry the steady-state
abundance of $\hhhp$ is straightforward to calculate and leads to a simple
and noteworthy result. Production of $\hhhp$ is via reaction (1.2), where
the cosmic ray ionisation rate, $\zeta$, is thought to be
$\sim$3~$\times$~10$^{-17}$~sec$^{-1}$ (e.g., see McCall \textit{et al.}
1999). Destruction is via reactions of type (3); of these CO (if not
largely frozen out on grains when T~$<$~20~K) is the
dominant reactant (k$_{CO}$~=~1.8~$\times$~10$^{-9}$~sec$^{-1}$, Anicich
\& Huntress 1986), although the
reaction of $\hhhp$ with atomic oxygen
also is important. Despite a high reaction rate, dissociative
recombination on electrons is unlikely in dark clouds because of the 
very low electron concentrations. Then, equating the rates of formation
and destruction,
   \begin{equation}
   \zeta n(\hh) \approx k_{CO} n(\hhhp) n(CO),
   \end{equation} \noindent relating in one simple equation perhaps the
three most important molecules in astronomy. Using the result from models
of dark cloud chemistry that n(CO)/n($\hh$) is approximately constant at
1.5~$\times$~10$^{-4}$ (Lee \textit{et al.} 1996), this equation reduces to
   \begin{equation} 
   n(\hhhp) \approx 1 \times 10^{-4}~cm^{-3}. 
   \end{equation} 

That the number density of $\hhhp$ is \textit{constant} in dark clouds is
highly unusual; the number densities of other molecular constituents of
the cloud scale as the total density. The behavior of $\hhhp$ derives from
its rates of production and destruction both scaling with the first power
of the cloud density, whereas production rates for most other molecules
scale as density squared.

Thus, the fractional abundance of $\hhhp$, n($\hhhp$)/n($\hh$) $\approx$
10$^{-4}$/n($\hh$), varies inversely with cloud density. If the cloud
density is known, 10$^{-4}$/n($\hh$) can be compared with
N($\hhhp)$/N($\hh$), where N($\hh$) has been measured or estimated (e.g.
from the silicate feature). In general cloud densities are not accurately
determined; however, studies of the collisional excitation of various
molecular species imply that densities in the W33A and GL2136 clouds, as
well as in most other clouds where $\hhhp$ has been sought, are
10$^{4}$~--~10$^{5}$~cm$^{-3}$. An additional uncertainty is that
densities probably vary significantly within these clouds.  For the above
density range n($\hhhp)$/n($\hh$) is 10$^{-8}$~--~10$^{-9}$. The values of
2~$\times$~10$^{-9}$ for N($\hhhp)$/N($\hh$) found toward W33A and GL2136
fall within this range, as do the values toward other sources for which
$\hhhp$ has been detected (McCall \textit{et al.} 1999). For clouds where
only upper limits are available for N($\hhhp$), the upper limits on
N($\hhhp)$/N($\hh$) are greater than 10$^{-9}$, and hence predicted values
of n($\hhhp)$/n($\hh$) are not ruled out.

A second test is to use the observed column densities of $\hhhp$ to
calculate the lengths of the absorbing columns of the clouds via the
relation L~$\approx$~N($\hhhp$)/n($\hhhp$) $\approx$ 10$^{4}$N($\hhhp$).  
For typical column densities (e.g., 5~$\times$~10$^{14}$~cm$^{-2}$) the
derived lengths are 1-2~pc (McCall \textit{et al.} 1999), where
1~pc~=~3.086~$\times~$10$^{18}$~cm. These are comparable to the measured
linear extents of the clouds on the sky, as would be expected. Where upper
limits to N($\hhhp$), and hence L, have been found, the results can be
reasonably explained by shorter absorption columns and/or denser clouds.

Thus, although only crude checks can be made at present, the derived
abundances of $\hhhp$ in dark clouds confirm the importance of cosmic-ray
induced ion-molecule chemistry in those environments.

\begin{figure}
\epsfxsize=10cm
\centerline{\epsfbox{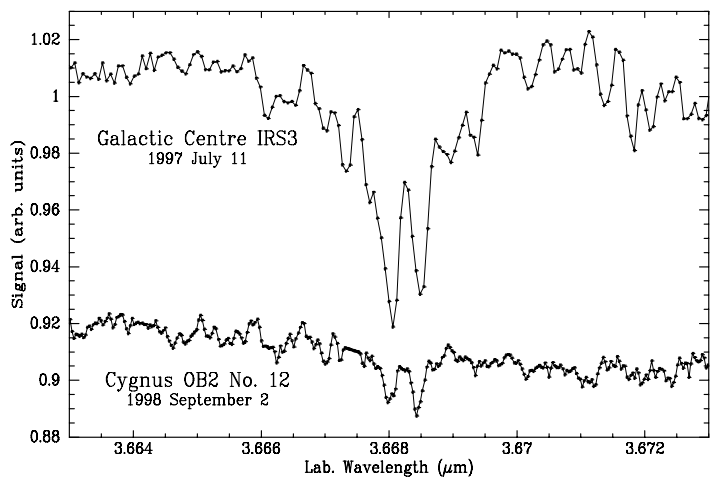}}
\caption{Spectra of the $\hhhp$ doublet toward the Galactic Centre source IRS3
at a resolution of 15 km s$^{-1}$ and the reddened star Cygnus OB2
No. 12 at a resolution of 9 km s$^{-1}$. $\hhhp$ in diffuse
clouds causes the narrow features seen towards the Cygnus source and
probably also towards IRS3.}
\end{figure}

\section{$\hhhp$ in diffuse interstellar gas}

In the course of the aforementioned survey of dark clouds for $\hhhp$ by
McCall \textit{et al.} (1999), strong absorption by the 3.67~$\mu$m doublet
was discovered along lines of sight to two infrared sources in the Galactic
centre on 1997 July 11 (Geballe \textit{et al.} 1999; figure~5). The visual
extinction toward the Galactic centre, $\sim$30~mag (Geballe \textit{et
al.} 1989),
is comparable to those
toward infrared sources in dark clouds in which $\hhhp$ had been found, but
N($\hhhp$) is nearly an order of magnitude greater. The line of sight to
the Galactic center, some 8~kpc long, is known to contain both dark clouds
and diffuse (low density) clouds which are penetrable at least to some
extent by visible radiation. (McFadzean \textit{et al.} 1989). Thus
interpretation of the spectra is not obvious, although it is clear from the
velocity profile of the absorption that some of the $\hhhp$ is located in
dark clouds known from radio/millimetre and infrared spectroscopy.

The previously unforeseen possibility that $\hhhp$ in low density clouds
also could be detectable prompted a search (on the same night) for $\hhhp$
toward one of the archetypal probes of the diffuse interstellar medium, the
visible star Cygnus OB2 No. 12, which is obscured by 10 visual magnitudes.
The $\hhhp$ doublet was easily detected there (figure 5). Analysis of this
spectrum and one obtained later containing an additional line yields
N($\hhhp$)~=~3.8~$\times$~10$^{14}$~cm$^{-2}$, comparable to that in dark
clouds, and T$\sim$30~K (McCall \textit{et al.} 1998).

Understanding this result has proved more elusive than understanding the
observations of $\hhhp$ in dark clouds. In a classical diffuse cloud H and
$\hh$ are roughly equally abundant and there is little CO (van Dishoeck \&
Black 1986). Little H is ionised by ultraviolet photons except at the edges
of the cloud. The $\hh$ within the cloud also is shielded from ionisation,
which requires photons of $\geq$15.4~eV, by the ionisation of atomic
hydrogen (requiring $\geq$13.6~eV photons) on the periphery. Within the
cloud $\hhhp$ is formed in the same way as in dense clouds (reaction 1.2
followed by 1.1). Destruction of $\hhhp$, however, is expected to be
dominated by electron recombination, because the essentially complete
single ionisation of gaseous atomic carbon produces a much higher
concentration of electrons than in dark clouds. (Despite depletion onto
dust particles, atomic carbon is abundant in diffuse clouds and requires
only $\geq$11.3~eV photons for ionisation to C$^{+}$.) Cardelli \textit{et
al.} (1996) and Sofia \textit{et al.} (1997) have measured the abundance
ratio of atomic carbon to hydrogen, z$_{C}$, to be 1.4~$\times$10$^{-4}$ in
diffuse clouds.  As in the case of dark clouds, the density of the
$\hhhp$-destroyer (in this case electrons, rather than CO), roughly scales
with the hydrogen density. Thus, a simple expression again is obtained for
the steady state density of $\hhhp$ (Geballe \textit{et al.} 1999),
   \begin{equation}
n(\hhhp) \approx \zeta /(4k_{e}z_{C}) \approx 1 \times 10^{-7} cm^{-3},
   \end{equation}
\noindent where
k$_{e}$~=~2.1~$\times$~10$^{-6}~$T$^{-0.5}$~cm$^{3}$~s$^{-1}$
(Sundstr\"{o}m \textit{et al.} 1998) is evaluated at 30~K. Once again the
density of $\hhhp$ is roughly constant, but in a diffuse cloud its value
is roughly three orders of magnitude less than in a dark cloud. As the
total gas density in a typical diffuse cloud is also a few orders of
magnitude less, concentrations of $\hhhp$ are comparable in the two
environments.

If the absorbing low density cloud or collection of clouds between Cygnus
OB2 No. 12 and the earth are as described above, the aggregate absorption
path length must be $\sim$~1~kpc in order to produce the observed line
strengths. This length, roughly half the 1.7~kpc distance from the earth to
the Cygnus OB2 association (Torres-Dodgen \textit{et al.} 1991), seems
physically unreasonable. The mean gas density along such an absorbing path,
$\sim$10~cm$^{-3}$, is insufficient for $\hh$ to be relatively abundant and
hence for $\hhhp$ to form. It also is inconsistent with observations of
C$_{2}$ (Souza \& Lutz 1977; Gredel \& M\"{u}nch 1994), and CO (Geballe
\textit{et al.} 1999) toward Cygnus OB2 No. 12, which imply that the
C$_{2}$ is located in clouds with densities at least an order of magnitude
higher and that the CO exists at even higher densities. Different values
for $\zeta$ and/or k$_{e}$ could explain the discrepancy. The cosmic ray
ionisation rate may be larger in the vicinity of an association of hot
stars such as Cygnus OB2 than in an isolated dark cloud, but the difference
is unlikely to be an order of magnitude. Alternatively, at low temperatures
k$_{e}$ may have to be considerably smaller to bring the above model into
agreement with the observations.

Recently Cecchi-Pestellini \& Dalgarno (2000) have suggested that much of
the material obscuring Cygnus OB2 No. 12 is in clumps of gas containing
both dense and diffuse components.  Using a somewhat higher cosmic ray
ionisation rate than adopted here, they fit the observations of $\hhhp$,
C$_{2}$, and CO with nine such cloudlets of density 10$^{2}$~cm$^{-3}$,
embedded in some of which are much higher density cores where most of the
carbon is in the form of CO.  The summed column length through these clumps
is $\sim$60~pc, far less than the derived $\hhhp$ absorption length of
1~kpc for classical diffuse clouds. High resolution spectroscopy of C$_{2}$
(Gredel \& M\"{u}nch 1994) shows the presence of four distinct low density
cloudlets along the line of sight to Cygnus OB2 No. 12, while millimetre
and infrared spectroscopy of CO (Geballe \textit{et al.} 1999) demonstrates
that at least two of these contain CO and that some of the CO probably is
located in dense regions. More sensitive measurements may reveal additional
clumps.

\section{Conclusion}

The observations to date demonstrate that, as predicted, $\hhhp$ is
an ubiquitous constituent of dark clouds.  The detected column densities
and upper limits are consistent with its production by cosmic ray
ionisation of $\hh$ and destruction via reactions with neutrals which form
the base of an extensive ion-molecule reaction network. Thus the
detection of $\hhhp$ provides a crowning confirmation of the theories of
Herbst \& Klemperer and Watson, proposed nearly three decades ago to
account for the rich chemistry observed in these clouds.

Because its density is constant in dark clouds, $\hhhp$ is a unique tool
for astronomers, with the potential of determining two fundamental
parameters: line of sight distances in the clouds and accurate values of
$\zeta$, the cosmic ray ionisation rate.  However, a glance at equation 4.1
reveals that a measurement of N($\hhhp$) only determines the product of
$\zeta$ and the column length.  More sophisticated approaches are needed to
determine the values of these fundamental parameters.  These could involve,
for example, more detailed modelling of well-observed dark clouds to
determine line of sight distances and density profiles. Direct measurements
of the column densities of $\hh$ and CO can provide additional constraints.
Alternatively, statistical studies of $\hhhp$ in many clouds, using
background infrared sources could provide multiple line of sight distances
through each cloud, to be compared with the cloud's linear extent on the
sky.  Whatever strategies are employed, observational progress clearly
requires the use of the new and future generations of large telescopes and
sensitive spectrometers, as many of the background sources will be
considerably fainter than those that have been utilised to date to detect
$\hhhp$.

The surprisingly large amounts of $\hhhp$ found toward the Galactic centre
and especially toward Cygnus OB2 No. 12 suggest that our understanding of
the physical conditions of the gas on these sight lines needs refinement.
The model of Cecchi-Pestellini and Dalgarno (2000) for the material in
front of Cygnus OB2 reminds us that environments intermediate between
classical dark and classical diffuse clouds can exist. It also indicates
that spectroscopy of $\hhhp$, whose steady state abundance is highly
sensitive to the densities of the neutrals, $\hh$ and CO, and the
electrons, can play a key role in characterising these environments.
Measurements of $\hhhp$, CO and other molecules toward additional obscured
stars in Cygnus OB2 along with searches for $\hhhp$ in additional objects
obscured by diffuse gas are important next steps towards understanding
these largely low density environments.

The discovery of $\hhhp$ toward the Galactic centre immediately suggests
the possibility of detecting $\hhhp$ in extragalactic environments. An
initial search on UKIRT is already under way, but it is clear that
observations using 8-10m telescopes will be required to reach more
than a few of the promising candidate galaxies. In the longer term, with
telescopes such as the Next Generation Space Telescope and even larger
aperture ground-based telescopes, we can anticipate that spectroscopy of
$\hhhp$ in external galaxies, in combination with observations of CO and
other molecules, will be a standard technique used to probe in detail the
properties of interstellar gas in the distant universe.

\begin{acknowledgements}

I wish to thank T. Oka and B. J. McCall, not only for helpful comments
during the preparation of this paper, but also for many enlightening
discussions during our years of collaboration. I am especially
grateful to the staff, present and past, of the United Kingdom Infrared
Telescope and indeed to all of those who have supported it.

\end{acknowledgements}

\label{lastpage}

\end{document}